\documentclass{PoS}

\usepackage{amsmath,amssymb,multirow}

\usepackage[sort&compress,numbers,colon]{natbib}

\usepackage{tikz}
\usetikzlibrary{positioning,arrows,patterns}
\usetikzlibrary{decorations.pathmorphing}
\usetikzlibrary{decorations.markings}
\usetikzlibrary{calc}
\tikzset{
    photon/.style={decorate, decoration={snake}, draw=black, thick},
    fermionnoarrow/.style={draw=black, postaction={decorate}, thick},
    scalar/.style={draw=black, postaction={decorate}, thick, dashed},
    fermion/.style={draw=black, postaction={decorate},decoration={markings,mark=at position .55 with {\arrow{>}}}, thick},
    gluon/.style={decorate, draw=black, decoration={coil,amplitude=4pt, segment length=5pt}, thick},
    vertex/.style={draw,shape=circle,fill=black,minimum size=3pt,inner sep=0pt} 
}

\newcommand{\lag}{\mathcal{L}}
\newcommand{\lhp}{\lambda_{h\phi}}
\newcommand{\set}[1]{\mathbb{#1}}
\newcommand{\eg}{\emph{e.g.}}
\newcommand{\sigv}{\langle \sigma v \rangle}

\newcommand{\modeqref}[1]{Eq.~\eqref{#1}}
\newcommand{\tabref}[1]{Table~\ref{#1}}
\newcommand{\secref}[1]{Section~\ref{#1}}

\newcommand{\refcite}[1]{Ref.~\cite{#1}}

\newcommand{\figref}[1]{Figure~\ref{#1}}
\newcommand{\figsref}[1]{Figures~\ref{#1}}

\title{Semi-Annihilating Wino-Like Dark Matter}

\ShortTitle{Semi-Annihilating Wino-Like Dark Matter}

\author{\speaker{A. Spray}\\%
        University of Melbourne\\
        E-mail: \email{andrew.spray@coepp.org.au}}

\author{Y. Cai\\
        University of Melbourne\\
        E-mail: \email{yi.cai@unimelb.edu.au}}

\abstract{Semi-annihilation is a generic feature of dark matter theories with symmetries larger than $\set{Z}_2$.  We explore a model based on a $\set{Z}_4$-symmetric dark sector comprised of a scalar singlet and a ``wino''-like fermion $SU(2)_L$ triplet.  This is the minimal example of semi-annihilation with a gauge-charged fermion.  We study the interplay of the Sommerfeld effect in both annihilation and semi-annihilation channels.  The modifications to the relic density allow otherwise-forbidden regions of parameter space and can substantially weaken indirect detection constraints.  We perform a parameter scan and find that the entire region where the model comprises all the observed dark matter is accessible to current and planned direct and indirect searches.}

\FullConference{18th International Conference From the Planck Scale to the Electroweak Scale \\
		 25-29 May 2015\\
		 Ioannina, Greece }

\begin{document}

\section{Introduction}

The dark matter (DM) problem remains one of the most important questions in contemporary particle physics.  Measurements across a range of scales, from galaxy rotation curves to fluctuations in the cosmic microwave background, all point to the existence of a cold non-baryonic component of matter in the Universe.  However, no unambiguous non-gravitational signal has been found and the microscopic properties of DM remain unknown.  For this reason, it is important to consider as varied a range of DM phenomenology as possible, to check the effectiveness of planned searches.

\begin{figure}
  \centering
  \begin{tikzpicture}[node distance=1cm and 1.75cm]
    \coordinate (v1);
    \coordinate[above left = of v1, label=above left:$\chi$] (i1);
    \coordinate[below left = of v1, label=below left:$\chi$] (i2);
    \coordinate[above right = of v1, label=above right:{$V$}] (o1);
    \coordinate[below right = of v1, label=below right:{$V$}] (o2);
    \draw[fermionnoarrow] (i1) -- (v1);
    \draw[fermionnoarrow] (v1) -- (i2);
    \draw[fermionnoarrow] (o2) -- (v1);
    \draw[fermionnoarrow] (v1) -- (o1);
    \draw[fill = white] (v1) circle (1);
    \fill[pattern = north west lines] (v1) circle (1);
  \end{tikzpicture}\quad
  \begin{tikzpicture}[node distance=1cm and 1.75cm]
    \coordinate (v1);
    \coordinate[above left = of v1, label=above left:$\chi_1$] (i1);
    \coordinate[below left = of v1, label=below left:$\chi_2$] (i2);
    \coordinate[above right = of v1, label=above right:{$\chi_3$}] (o1);
    \coordinate[below right = of v1, label=below right:{$V$}] (o2);
    \draw[fermionnoarrow] (i1) -- (v1);
    \draw[fermionnoarrow] (v1) -- (i2);
    \draw[fermionnoarrow] (o2) -- (v1);
    \draw[fermionnoarrow] (v1) -- (o1);
    \draw[fill = white] (v1) circle (1);
    \fill[pattern = north west lines] (v1) circle (1);
  \end{tikzpicture}\quad
  \begin{tikzpicture}[node distance=1cm and 1.75cm]
    \coordinate (v1);
    \coordinate[above left = of v1, label=above left:$\chi_1$] (i1);
    \coordinate[below left = of v1, label=below left:$\chi_2$] (i2);
    \coordinate[above right = of v1, label=above right:{$\chi_3$}] (o1);
    \coordinate[below right = of v1, label=below right:{$\chi_4$}] (o2);
    \draw[fermionnoarrow] (i1) -- (v1);
    \draw[fermionnoarrow] (v1) -- (i2);
    \draw[fermionnoarrow] (o2) -- (v1);
    \draw[fermionnoarrow] (v1) -- (o1);
    \draw[fill = white] (v1) circle (1);
    \fill[pattern = north west lines] (v1) circle (1);
  \end{tikzpicture}
  \caption{Three types of dark sector processes, where $\chi$ ($V$) is a dark (visible) sector field.  (Left): DM annihilation to/from the SM; this is the only process possible when the dark matter is stabilised by a $\set{Z}_2$ symmetry.  (Centre): Semi-annihilation, a non-decay process with an odd number of external visible particles.  (Right): DM exchange, only possible when the dark sector is multicomponent.}\label{fig:SAandDE}
\end{figure}
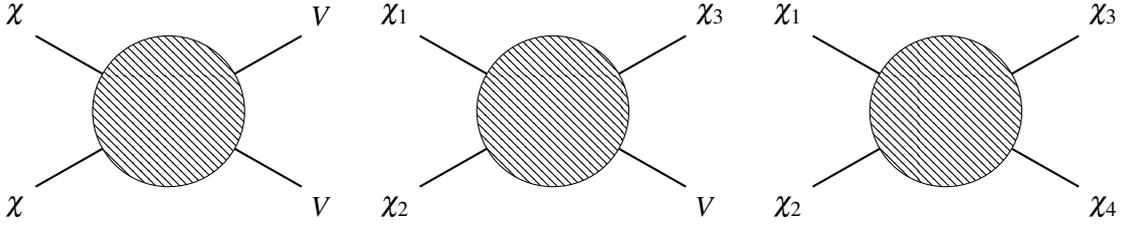

Semi-annihilation (SA) is a generic feature of dark sector phenomenology that occurs whenever the symmetry that stabilizes DM is larger than $\set{Z}_2$~\cite{D'Eramo:2010ep}.  We illustrate it in \figref{fig:SAandDE}.  For the usually-considered case, the only allowed $2\to 2$ diagram is that on the left: DM annihilation to/from or scattering off the SM.  SA is shown by the central diagram, and is characterised by a non-decay process with an odd number of external dark sector particles.  Finally, many models of semi-annihilating dark matter (SADM) involve multicomponent dark sectors, in which case dark matter exchange (DME), the process shown in the right diagram, can be relevant.

Previous studies of SADM have mostly focused on scalar or vector DM, see \eg~\cite{D'Eramo:2010ep,D'Eramo:2012rr,Belanger:2012vp,Ivanov:2012hc,Belanger:2014bga} (but see \refcite{Aoki:2014cja} for an exception).  This is because bosons can have renormalisable SA self-couplings, \emph{e.g.} a cubic term in a $\set{Z}_3$ theory.  Equivalently, fermionic examples of SADM \emph{necessitate} multi-component dark sectors, enriching the phenomenology.  The dark sector must include bosons, to avoid an accidental $\set{Z}_2$ forbidding fermionic SA.  Thus the minimal model consists of a Dirac fermion $\psi$ and a real scalar singlet $\phi$, with respective charges 1 and 2 under a $\set{Z}_4$ symmetry.\footnote{A $\set{Z}_3$ symmetry would require a complex scalar singlet and would introduce additional interaction terms.}  Unfortunately, $\psi$ has no direct couplings to the SM which impedes the observation of fermionic SA.  This leads us to consider a next-to-minimal ``wino-like'' model, where $\psi$ is an $SU(2)_L$ triplet with zero hypercharge.  There are three physical Dirac fermion states, two charged ($\psi^\pm$) and one neutral ($\psi^0$); loop effects split them by $\delta m_\psi \equiv m_{\psi^\pm} - m_{\psi^0} \approx 167$~MeV~\cite{Cirelli:2005uq}.  Analogously with the wino, we expect the Sommerfeld effect (SE) to be important when $m_\psi \sim m_W/\alpha_2 \sim 2.5$~TeV.  This is a non-perturbative effect at low velocities due to long-range interactions, distorting the two-particle wave function at the interaction point, and can enhance cross sections by orders of magnitude~\cite{Hisano:2004ds,Hisano:2006nn}.

\begin{table}
  \centering
  \begin{tabular}{|c|c|c|c|}
    \hline
    Field &Type &  $G_{SM}$ & $\set{Z}_4$ \\
    \hline
    $\phi$ & Real Scalar & (1, 1, 0) & 2 \\
    $\psi \sim (\psi^+, \psi^0, \psi^-)$ & Dirac Fermion & (1, 3, 0) & 1 \\
    \hline
  \end{tabular}
  \caption{New particle content for the model we consider in this paper.}\label{tab:parts}
\end{table}

We summarise the new particle content in \tabref{tab:parts}.  The Lagrangian for this theory is 
\begin{align}
  \lag & = \lag_{SM} + \bar{\psi} (i D\!\!\!\!\!/ - m_\psi) \psi + \frac{1}{2} (\partial_\mu \phi)^2 + \frac{1}{2} (m_\phi^2 \phi^2 - \lhp v^2) \notag \\
  & \quad + (y \phi \, \bar{\psi}^c \psi + h.c. ) + \frac{1}{2} \, \lhp \, H^\dagger H \, \phi^2 + \frac{1}{4} \, \lambda_{4\phi} \phi^4 \,.
\end{align}
There are five new parameters compared to the SM: the masses of the two dark sector particles $m_\psi$ and $m_\phi$, the Higgs portal coupling $\lhp$, the semi-annihilation coupling $y$ and the new scalar quartic $\lambda_{4\phi}$.  The last of these is phenomenologically unimportant, so we effectively have a four-dimensional parameter space.  We may take $y$ real and positive without loss of generality.  In the limit when $y \to 0$, this model reduces to the combination of a Dirac wino-like fermion and of a scalar singlet coupled through a Higgs portal~\cite{Cline:2013gha}.  In these proceedings we discuss our study of this model from \refcite{Cai:2015zza}.  In \secref{sec:SASE}, we consider the interplay of SA and the SE on the relic density and indirect detection.  We then apply those results in \secref{sec:f3} to a scan of the fermion triplet parameter space.

\section{Semi-Annihilation and the Sommerfeld Effect}\label{sec:SASE}

The two aspects of DM phenomenology directly affected by SA and the SE are the relic density and indirect detection.  Both involve processes initiated by two dark sector states (so the SE is important) and potentially with dark sector final states (so SA is relevant).  Other aspects of phenomenology, \eg\ production at colliders or direct detection, are either unaffected or only indirectly sensitive through the modification to the relic density.

The computation of the relic density in general SA models may be found in \eg\ \refcite{Belanger:2014vza}.  For our specific model, the coupled Boltzmann equations take the form
\begin{align}
  \frac{d Y_\Psi}{d x} & = \frac{s Z}{H x} \, \biggl[ \bigl(Y_\Psi^2 - (Y_\Psi^{eq})^2 \bigr) \sigv (\Psi\Psi \to SM) + \biggl( Y_\Psi^2 - Y_\phi \frac{(Y_\Psi^{eq})^2}{Y_\phi^{eq}} \biggr) \sigv (\Psi\Psi \to\phi SM) \notag \\
  & \quad + \biggl( Y_\Psi^2 - Y_\phi^2 \frac{(Y_\Psi^{eq})^2}{(Y_\phi^{eq})^2} \biggr) \sigv (\Psi\Psi\to\phi\phi) \biggr] \,, \label{eq:fermdY} \displaybreak[0]\\
  \frac{d Y_\phi}{d x} & = \frac{s Z}{H x} \, \biggl[ \bigl(Y_\phi^2 - (Y_\phi^{eq})^2 \bigr) \sigv (\phi\phi \to SM) + Y_\Psi \bigl( Y_\phi - Y_\phi^{eq} \bigr) \sigv (\Psi\phi \to \Psi SM) \notag \\
  & \quad + \frac{1}{2} \biggl( Y_\phi \frac{(Y_\Psi^{eq})^2}{Y_\phi^{eq}} - Y_\Psi^2 \biggr) \sigv (\Psi\Psi \to \phi SM) + \biggl( Y_\phi^2 \frac{(Y_\Psi^{eq})^2}{(Y_\phi^{eq})^2} - Y_\Psi^2 \biggr)\sigv (\Psi\Psi \to \phi\phi) \biggr] \,. \label{eq:scadY}
\end{align}
Here, $s$ is the entropy density, $H$ the Hubble rate, $x = T_0/T$ the inverse temperature, $Y_i = n_i/s$ the abundance of species $i$, and $Z \approx 1$ is a function of the number of relativistic degrees of freedom.  $\Psi$ denotes a sum over all fermion and antifermion species.  Scattering off the SM maintains thermal equilibrium among the different fermion components.  This suppresses $Y_{\psi^\pm}$ when $T \lesssim \delta m_\psi$, affecting the SE which is relevant at these late times.  The second term of \modeqref{eq:fermdY}, and the second and third terms of \modeqref{eq:scadY}, are SA; while the last terms of both equations are DME.  The SE modifies all processes with two fermions in the initial state, specifically all terms in \modeqref{eq:fermdY} and the second line of \modeqref{eq:scadY}.  We approximate the SE as only applying to the $s$-wave piece of the (semi-)annihilation cross section, so that
\begin{equation}
  \sigv = \langle \mathcal{S} \rangle \, \sigma_0 + ( \sigv_0 - \sigma_0) \,,
\end{equation}
with $\sigv_0$ ($\sigma_0$) the unenhanced thermally averaged ($s$-wave) cross section and $\langle\mathcal{S}\rangle$ the thermally averaged SE factor.  

In our computation of the SE, we follow the formalism of \refcite{Cirelli:2007xd}.  We split the calculation into independent subgroups by the unbroken quantum numbers: charge $Q$, angular momentum $J = S$ and $\set{Z}_4$ charge $q$.   For each subspace, we solve a Schr\"odinger equation for a generally matrix-valued two-particle wavefunction $\Phi_{ij}$:
\begin{equation}
  - \frac{1}{M} \, \Phi_{ij}'' (r) + \sum_k V_{ik} (r) \Phi_{kj} (r) = K \Phi_{ij} (r) \,,
\label{eq:SomEffSchEq}\end{equation}
with $M$ the mass,  $K$ the centre of momentum frame kinetic energy at large separation, $r$ the separation and $V_{ij} (r)$ the long-range potential.  The indices label different two-particle states.  The wavefunction satisfies the boundary conditions
\begin{equation}
  \Phi_{ij} (0) = \delta_{ij} \quad\text{and}\quad \lim_{r\to\infty} \frac{\Phi_{ij}'(r)}{\Phi_{ij} (r)} = i \sqrt{M (K - V_{ii} (\infty)) } \text{ (no sum)} \,.
\label{eq:SEBC}\end{equation}
The enhancement matrix $A_{ij}$ is given by
\begin{equation}
  A_{ij} = \lim_{r\to\infty} \frac{\Phi_{ij} (r)}{\exp(i \Re \sqrt{M (K - V_{ii} (\infty))} r)} \,,
\end{equation}
such that $A_{ij} = \delta_{ij}$ in the absence of the SE.  The cross section for each two-body initial state is
\begin{equation}
  \sigma_i = c_i (A \cdot \Gamma \cdot A^\dagger)_{ii} \,,
\end{equation}
where $\Gamma$ are annihilation matrices and $c_i = 1$ (2) if the state $i$ contains distinct (identical) particles.  

\begin{figure}
  \centering
  \includegraphics[width=0.49\textwidth]{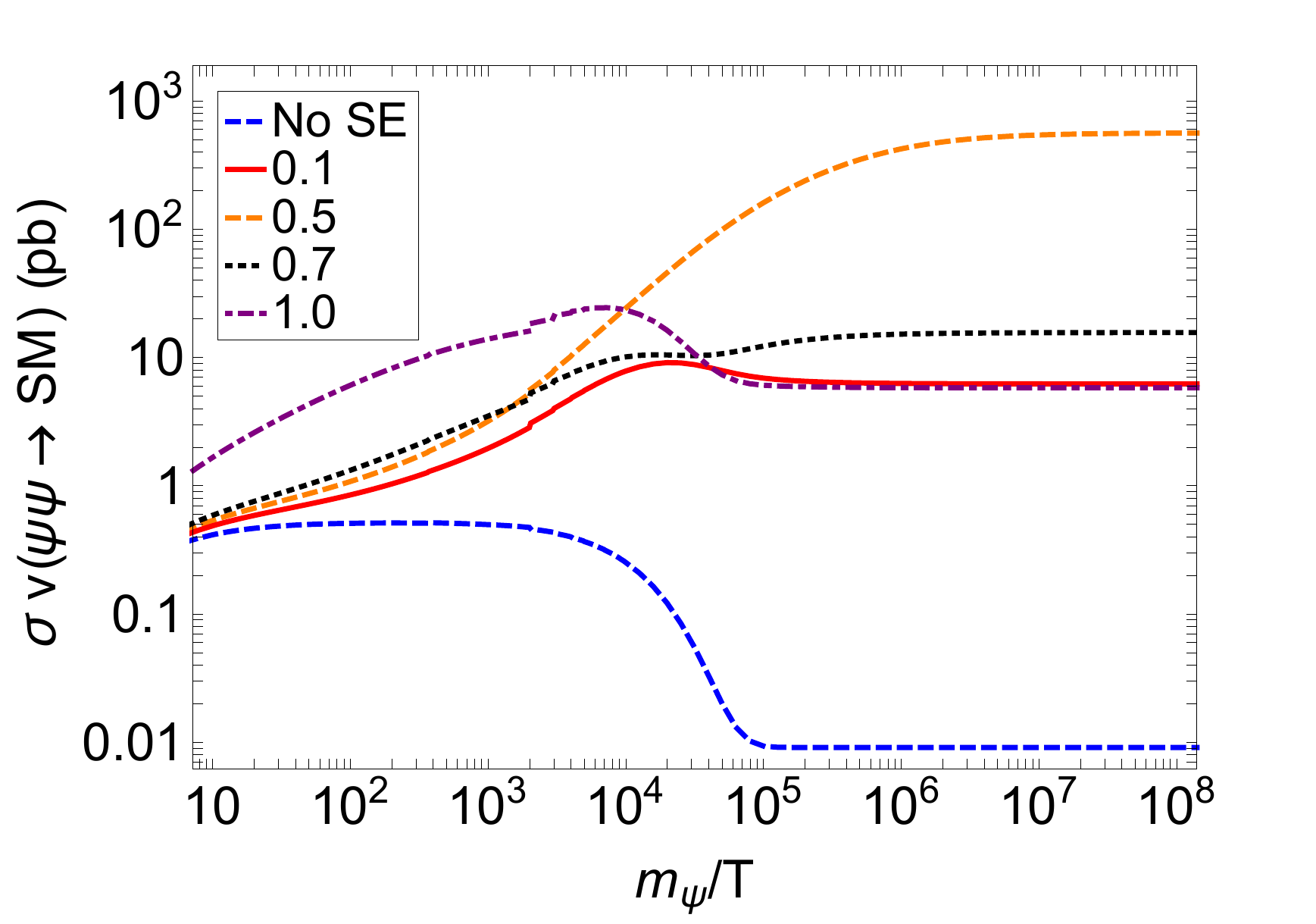}
  \includegraphics[width=0.49\textwidth]{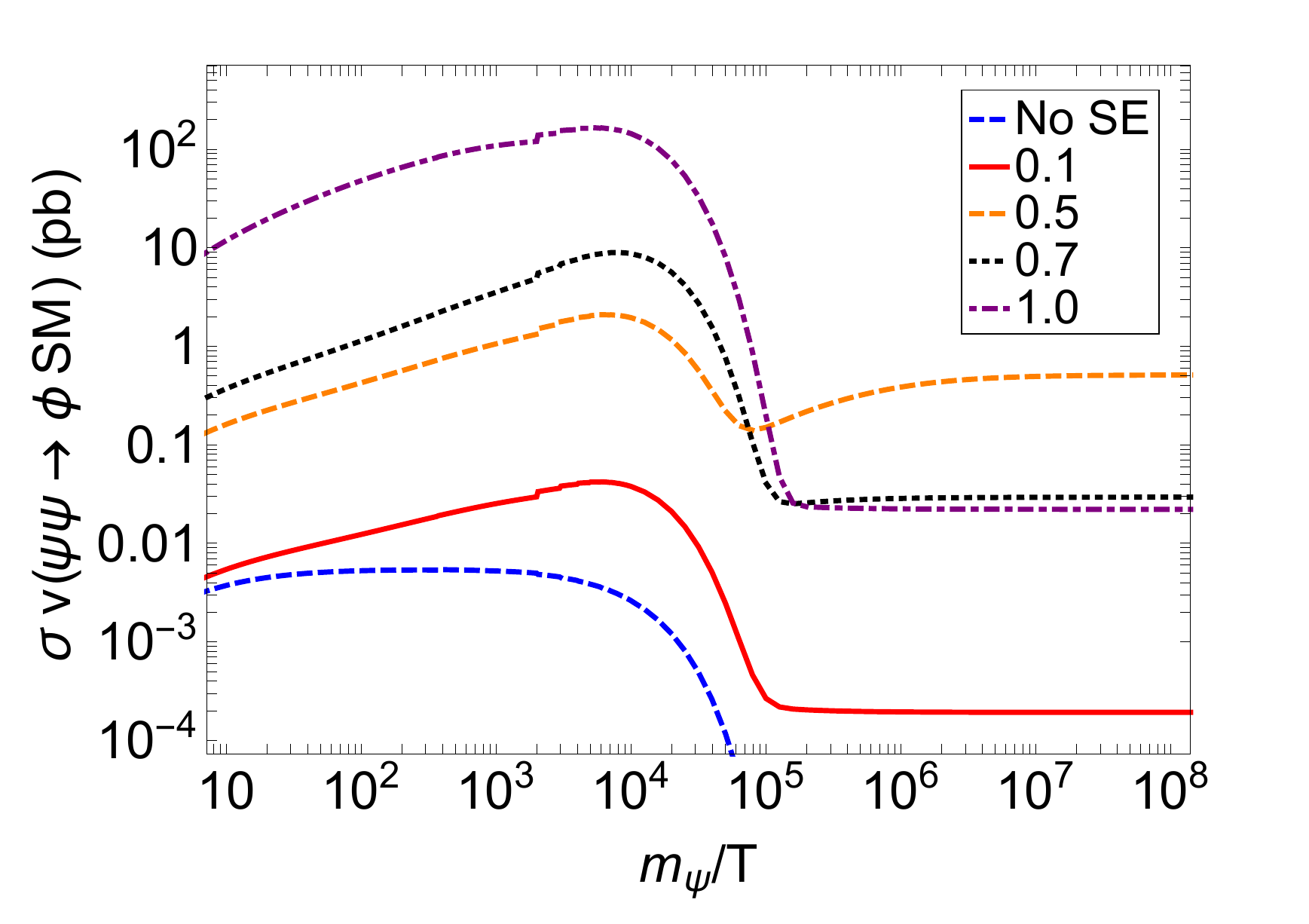}
  \caption{Thermally-averaged cross sections for annihilation (left) and SA (right).  These results are for $m_\psi = 2$~TeV, $m_\phi = 158$~GeV and values of $y$ as labelled, except for the blue dashed line which shows the cross sections without including the SE.}\label{fig:SE}
\end{figure}

The model-dependent elements of this calculation are the matrices $V$ and $\Gamma$.  Their diagonal entries have simple physical interpretations as the potential energies and annihilation cross sections of the associated two-body state.  The off-diagonal elements are more opaque, but the formal definitions are in terms of the real ($V$) and imaginary ($\Gamma$) parts of the generalised two-body propagator $(\Psi\Psi)^i\to (\Psi\Psi)^j$.  Explicit expressions for $V$ and $\Gamma$ in our model are given in \refcite{Cai:2015zza}.  We plot the thermally averaged annihilation and SA cross sections in \figref{fig:SE}.  We see that in the absence of the SE, SA effectively vanishes at low temperatures.  This is due to the previously discussed thermal suppression in the abundance of charged fermions when $T < \delta m_\psi$.  In contrast to $\psi^0\bar{\psi}^0 \to W^+W^-$, the $\psi^0\psi^0$ state can not semi-anihilate.  It follows that the SE is relatively more important to the SA channel than to annihilation.

The calculation of the SE for indirect detection follows a very similar methodology.  The main difference is that all charged fermions have decayed at late times.  This means that only the $Q = 0$ subspaces contribute, and we must modify the boundary conditions to enforce no charged states at infinite separation.  The fact that the $\psi^0\psi^0$ state does not interact means that indirect signals from SA are suppressed.  This is also true for $\gamma$-ray lines from SA despite the fact that both $\psi^0\bar{\psi}^0 \to \gamma\gamma$ and $\psi^0\psi^0 \to \phi \gamma$ have no tree-level contribution.  Both processes appear at one-loop, but this is not the leading contribution to the annihilation channel: instead, the SE-mediated process $\psi^0 \bar{\psi}^0 \to \psi^\pm \bar{\psi}^\mp \to \gamma\gamma$ dominates~\cite{Hryczuk:2011vi}.  The equivalent SA process $\psi^+ \psi^- \to \phi \gamma$ is a pure spin-1 process, but the $\psi^0\psi^0$ state can only exist in spin-0 by Fermi statistics.

\begin{figure}
  \centering
  \includegraphics[width=0.48\textwidth]{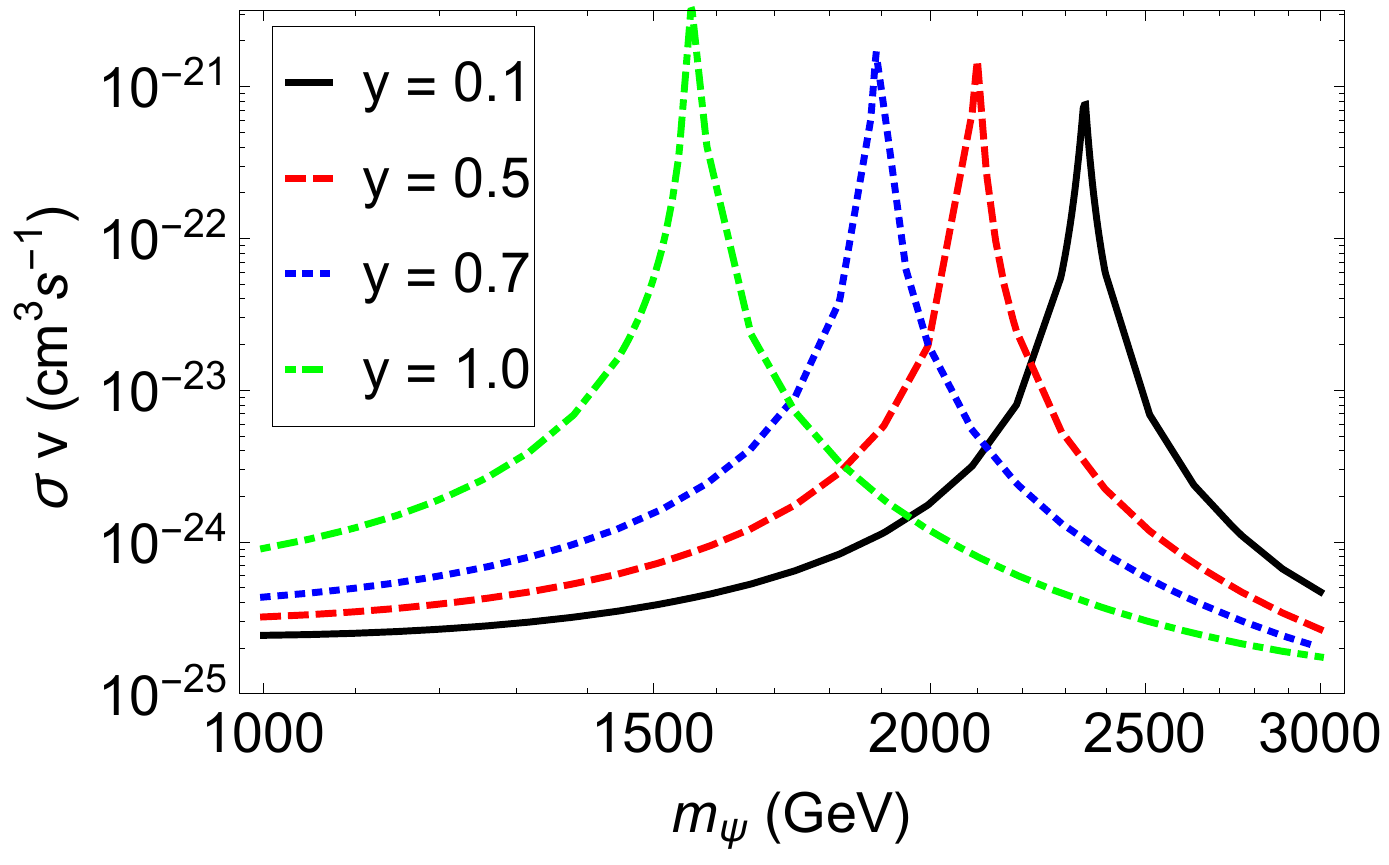}
  \includegraphics[width=0.48\textwidth]{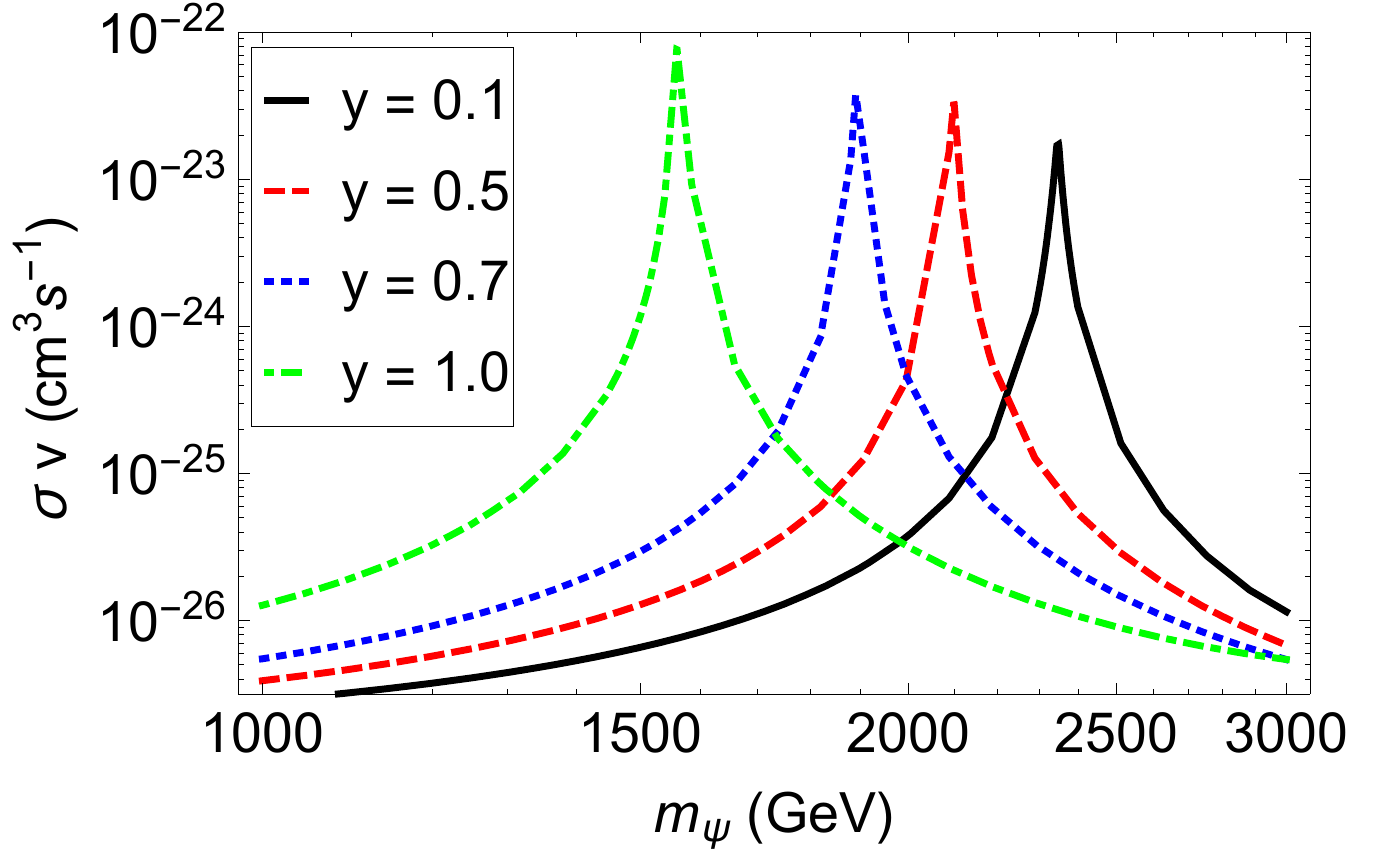}
  \caption{Combined annihilation cross sections to $WW$, $ZZ$ and $Z\gamma$ (left) and $\gamma\gamma$ and $Z\gamma$ (right), for $m_\phi = 200$~GeV and different values of $y$.  Note that the resonant peak occurs at smaller values of $m_\psi$ as $y$ increases.}\label{fig:IDXSec}
\end{figure}

\begin{figure}
  \centering
  \includegraphics[height=0.4\textwidth]{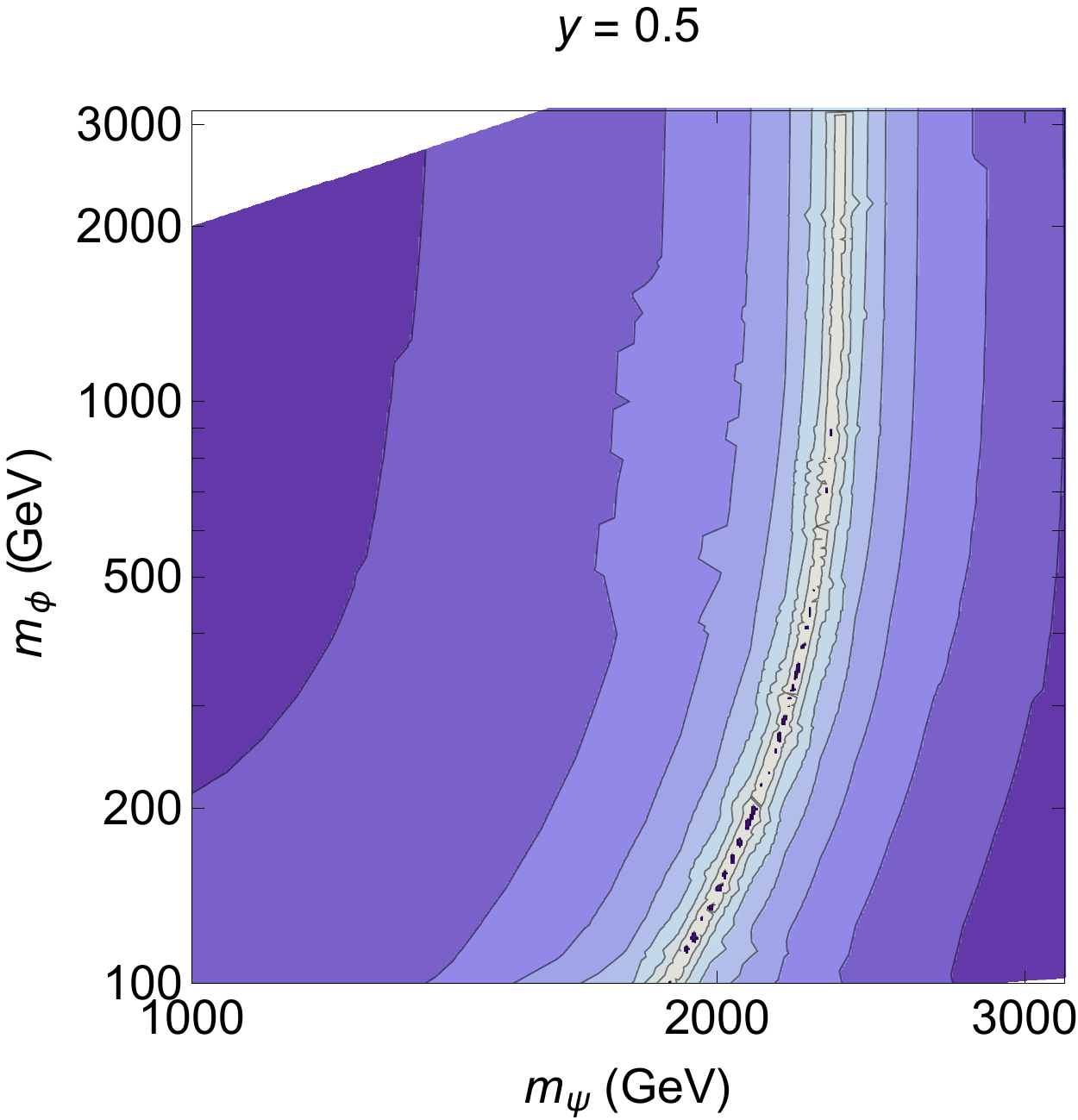}
  \includegraphics[height=0.4\textwidth]{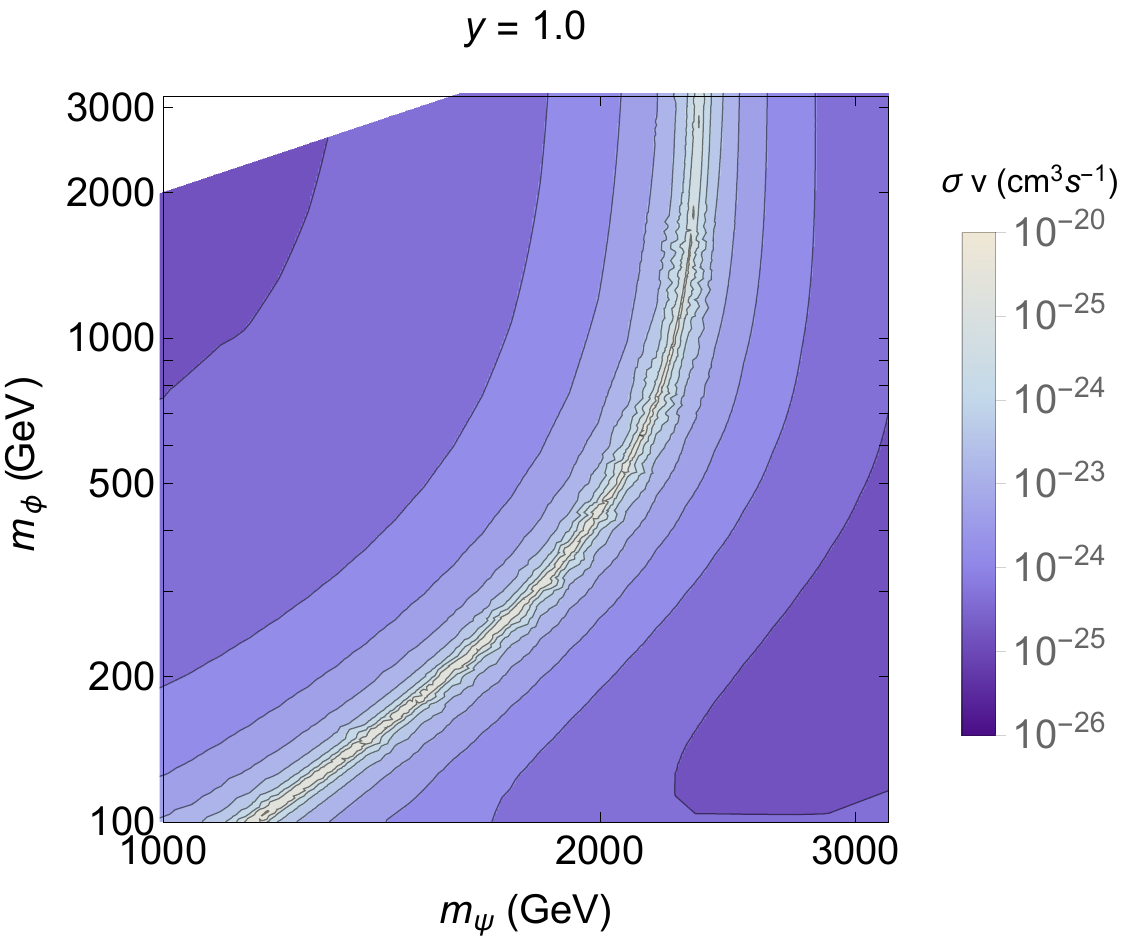}
  \caption{Annihilation cross sections for fermions to massive gauge bosons for $y = 0.5$ (1.0) in the left (right) figure.  Note that the resonant peak occurs at smaller values of $m_\psi$ as $m_\phi$ decreases, and this effect is stronger for larger $y$.}\label{fig:respos}
\end{figure}

However, indirect signals do retain some dependence on the coupling $y$.  When $m_\phi \ll m_\psi$, the scalar contributes to the potential, changing the position of the resonance.  We illustrate this in \figsref{fig:IDXSec} and~\ref{fig:respos}, where we plot the annihilation cross sections for different $y$ and $m_\phi$.  From \figref{fig:IDXSec}, we see that as $y$ increases the resonance becomes stronger and moves to smaller fermion masses, while \figref{fig:respos} demonstrates how this effect varies with the scalar mass.

\section{Fermion Triplet Phenomenology}\label{sec:f3}

We now use the results of \secref{sec:SASE} to explore the parameter space of the fermion triplet model.  We have a strict lower bound $m_\psi > 480$~GeV from LHC searches for disappearing tracks~\cite{Aad:2013yna,CMS:2014gxa}.  These constrain $pp \to W^\pm \to \psi^\pm\bar{\psi}^0$ followed by the displaced decay $\psi^\pm \to \psi^0 \pi^\pm$.  The upper bound on $m_\psi$ comes from requiring the total relic density be no more than observations;  we find $m_\psi \lesssim 5$~TeV.   A combination of the Higgs invisible width~\cite{Belanger:2013xza} and LUX observations~\cite{Akerib:2013tjd} then imply either 53~GeV$\,< m_\phi < 63$~GeV or $m_\phi > 130$~GeV.  We choose $\lhp = 0.1$ for which the region near the Higgs resonance is excluded.  While there is no absolute upper bound on the scalar mass, for $m_\phi > 2m_\psi$ the scalar is unstable and the phenomenology essentially reduces to that of a pure fermion triplet stabilised by a $\set{Z}_2$ symmetry.

\begin{figure}
  \centering
  \includegraphics[width=0.48\textwidth]{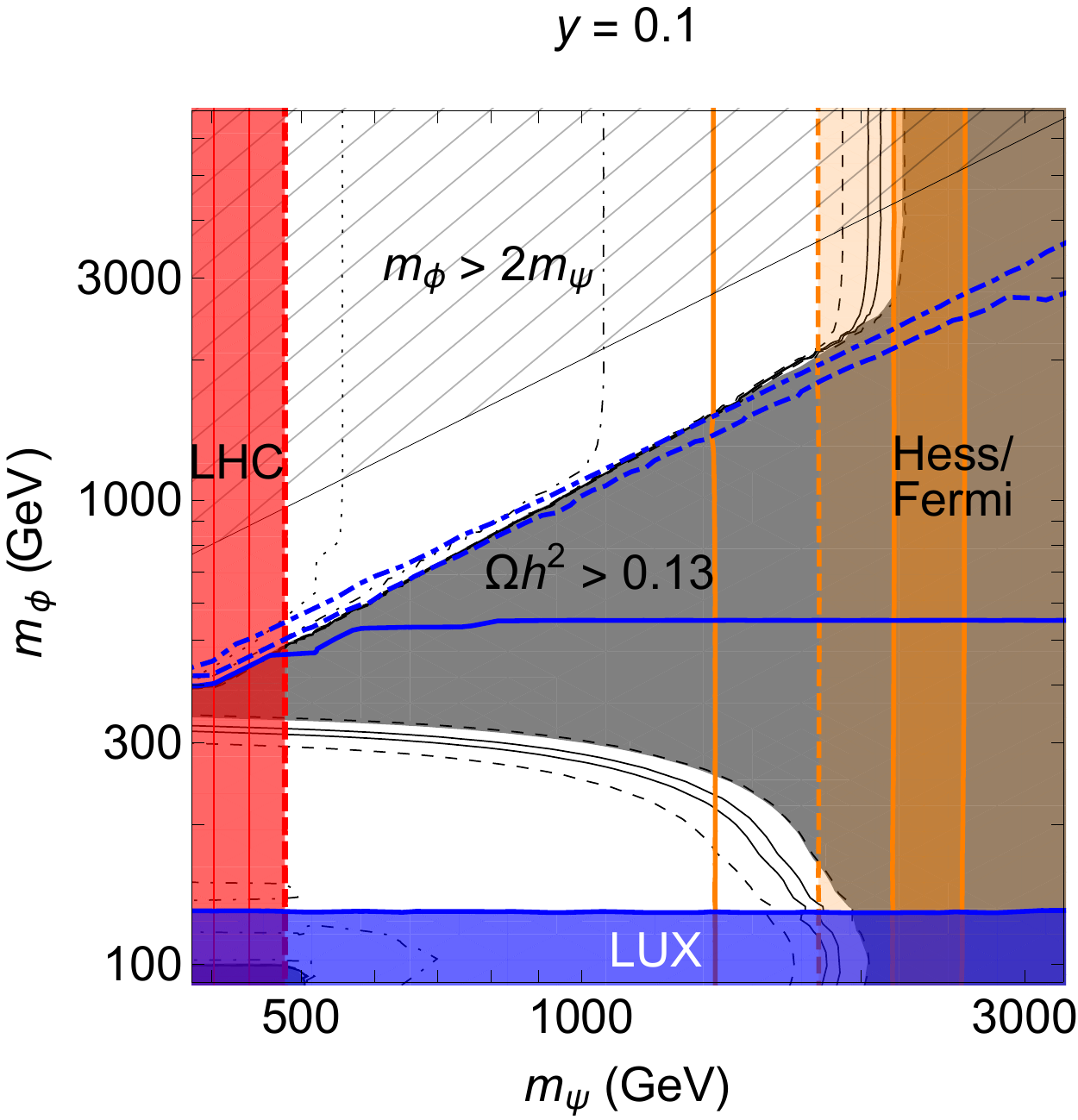}
  \includegraphics[width=0.48\textwidth]{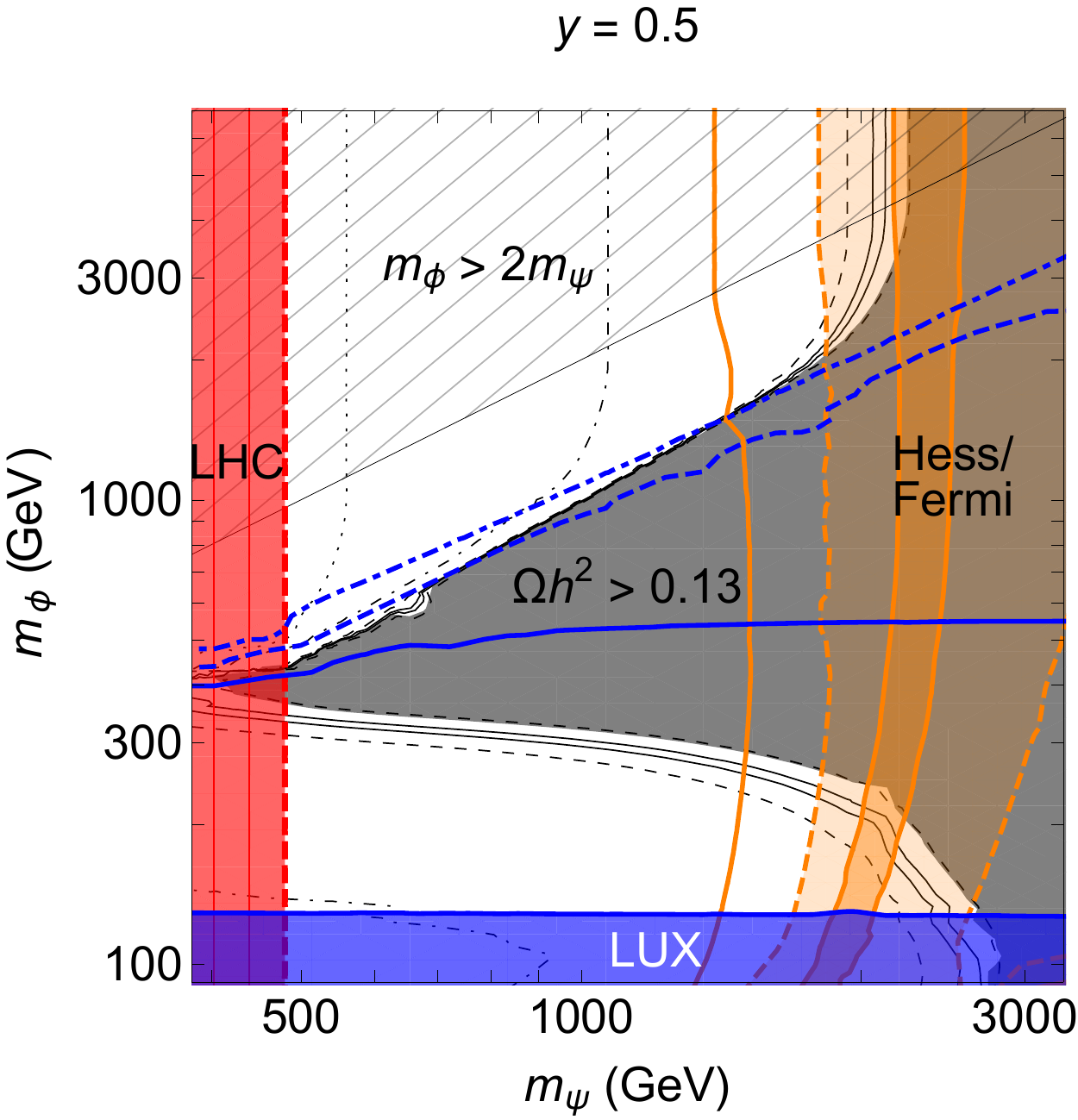}\\
  \includegraphics[width=0.48\textwidth]{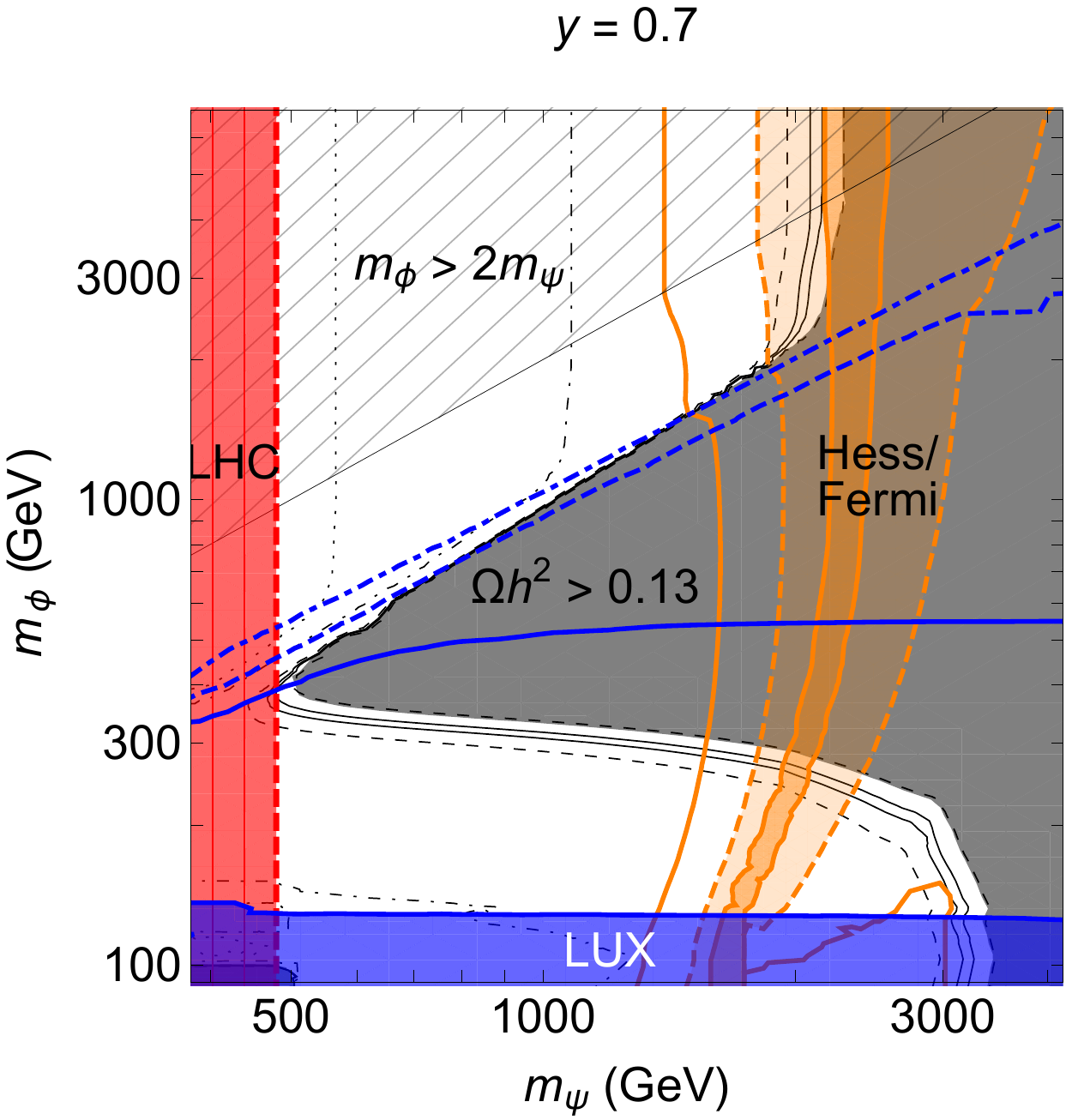}
  \includegraphics[width=0.48\textwidth]{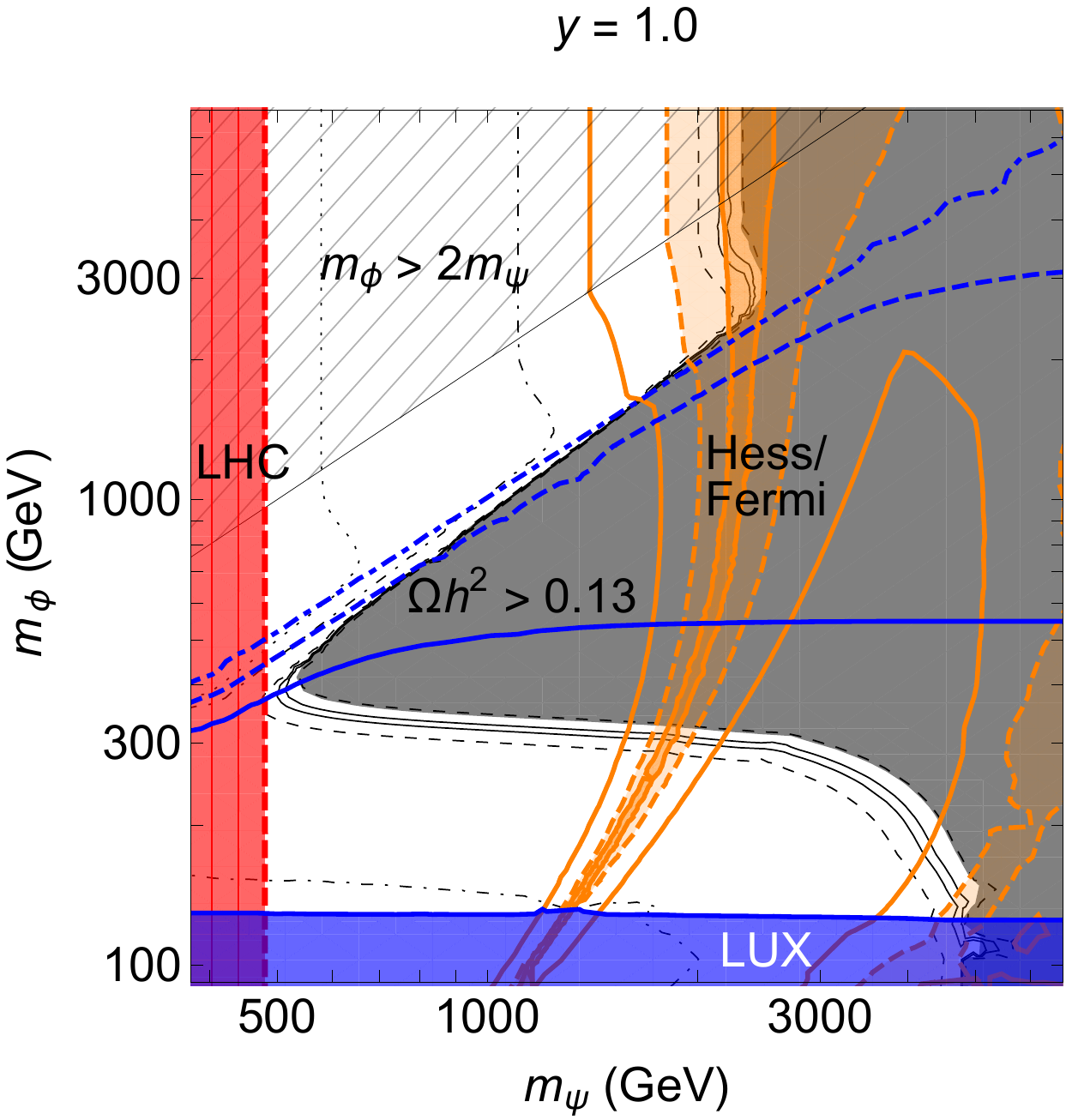}
  \caption{Slices of the fermion triplet parameter space in the $m_\phi$--$m_\psi$ plane, for $\lhp = 0.1$ and $y$ as labelled.  The grey shaded regions show regions where the total relic density is larger than observations.  In the white hatched region, the scalar decays to two fermions.  The red (blue, orange) shaded regions and contours show current and future bounds from the LHC (direct detection, indirect searches).  See the text for more details.}\label{fig:tripletresults}
\end{figure}

We scan the $m_\phi$--$m_\psi$ plane of parameter space for $y = 0.1$, 0.5, 0.7 and 1.0.  For $y = 0.1$, we expect SA to be subdominant to fermion and scalar annihilation and the two DM particles to freeze out independently.  Increasing $y$ will reveal the effect of SA.  We show our  results in in \figref{fig:tripletresults}.  The grey shaded region is excluded by a too large total DM relic density: $\Omega_{\phi + \psi} > 1.1 \,\Omega_{cdm}$, where we have allowed for a 10\% theoretical uncertainty.  Exclusions from LHC, LUX, HESS $\gamma$-ray lines from the galactic centre~\cite{Abramowski:2013ax} and Fermi diffuse photon fluxes~\cite{Ackermann:2015zua,Cirelli:2015bda} are as marked.  The HESS and Fermi limits are respectively the strongest limits for an (optimistic) NFW~\cite{Navarro:1995iw} and (conservative) cored DM density profile.  Prospective limits from LUX~\cite{Szydagis:2014xog} (Xenon1T~\cite{Aprile:2012zx}, LZ~\cite{Malling:2011va}) are shown by blue solid (dashed, dot-dashed) contours; and from CTA~\cite{Consortium:2010bc,Silverwood:2014yza}, assuming an optimistic profile, by the orange solid contour.  Note that all direct detection limits come from scalar scattering mediated by the Higgs portal; and for indirect detection from fermion annihilation to gauge bosons.

\begin{figure}
  \centering
  \includegraphics[height=0.32\textwidth]{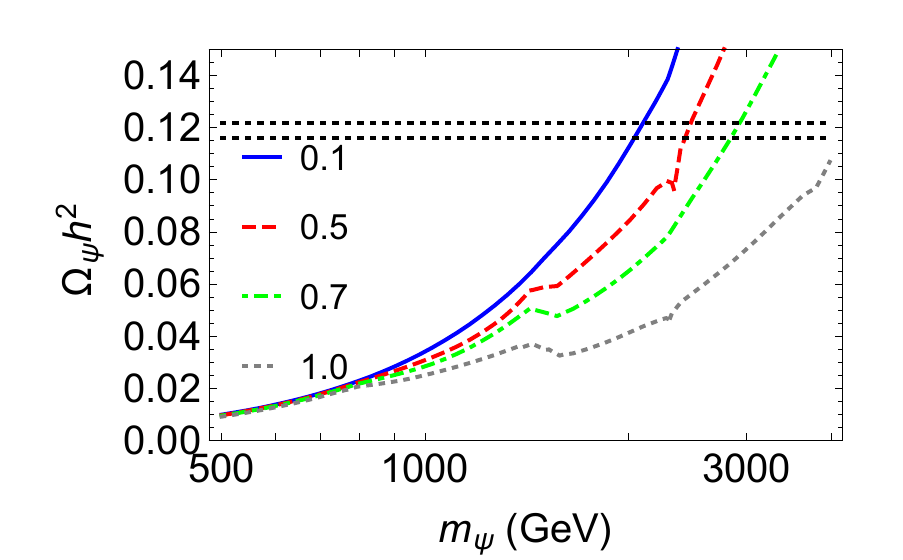}\hfill
  \includegraphics[height=0.32\textwidth]{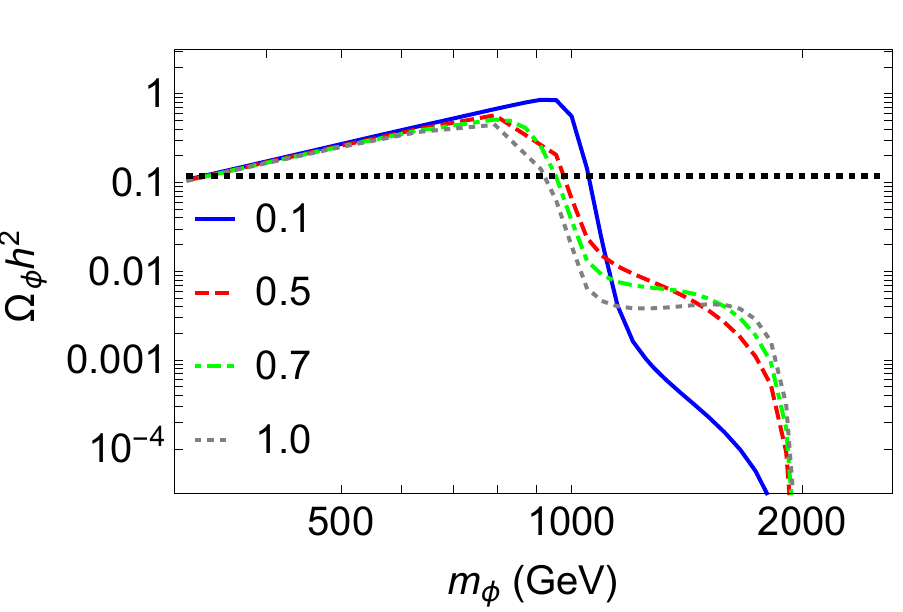}
  \caption{Fermion (left) and scalar (right) relic densities for different values of $y$ as labelled, and $m_\phi = 1.5$~TeV (left) or $m_\psi = 1$~TeV (right).  The dotted bands denote the Planck 3$\sigma$ measurement.}\label{fig:RDydep}
\end{figure}

We observe that for $y = 0.1$, $\Omega_\psi$ is a function only of $m_\psi$; while $\Omega_\phi$ is a function only of $m_\phi$, unless $m_\phi > m_\psi$.  This is as expected when SA is negligible; the only effect of $y$ comes from the DME process $\phi\phi \to \psi\bar{\psi}$ when $m_\phi > m_\psi$.  \figref{fig:RDydep} shows us how this changes as SA becomes important.  For the scalar at large $y$, we see the relic density drop for $m_\phi \lesssim m_\psi$, due to $\psi\phi \to \bar{\psi} + SM$; while it \emph{increases} for heavy scalars, due to $\psi\psi \to \phi + SM$.  The latter process is enhanced, so it can increase $\Omega_\phi$ by orders of magnitude.  For the fermion, when $m_\psi > m_\phi/2$ ($m_\phi$), SA (DME) depletes the relic density, though less dramatically as the annihilation is also enhanced.

This motivates splitting our analysis of \figref{fig:tripletresults} into two regions.  When $m_\phi < m_\psi$, the scalar freezes out independently of the fermion.  For small $y$, the fermion density is also set independently.  The correct relic density occurs when the two components accidentally sum to the observed value.  Due to the lower bounds on the DM masses, we find $\Omega_\phi \sim \Omega_\psi$ in this case.  For large $y$, SA and DME deplete the fermion abundance and shift the resonance as shown in \figref{fig:respos}.  This allows fermion masses $m_\psi \gtrsim 3$~TeV that are excluded for a pure wino.   Indirect constraints are much weaker and focused on the resonance.  However, we find that for all $y$, the full LUX data set can exclude this region as direct detection depends only on $\Omega_\phi$.

When $m_\phi \gtrsim m_\psi$, the correct relic density is typically produced for nearly equal masses.  This is due to the SA and DME-induced rapid variation in $\Omega_\phi$ visible in \figref{fig:RDydep}.  Along this line we find $\Omega_\phi \sim \Omega_\psi$.  The exception occurs when $m_\psi \approx 2.1$~TeV and fermion annihilation alone gives the correct relic density.  For sufficiently heavy scalars, SA and DME ensure that $\Omega_\psi \gg \Omega_\phi$ and the phenomenology mostly reduces to a pure fermion triplet.  The presence of SA and DME allow points in the $m_\phi$--$\lhp$ plane that are excluded for a pure scalar singlet.  The region in parameter space where our model produces the full DM relic density can be excluded for optimistic DM galactic profiles.  For $m_\psi \lesssim 1.5$~TeV, direct searches at LZ are most sensitive.  In the complementary region $m_\psi \gtrsim 1.5$~TeV, indirect searches are stronger.  As with a pure wino, HESS already excludes the region around $m_\psi \approx 2.1$~TeV and CTA can exclude the rest.  If the DM profile is more conservative, then exploring this part of parameter space would require a 100~TeV collider~\cite{Low:2014cba}.

Finally, let us briefly discuss the prospects for identifying this model if a discovery is made.  The best situation involves inconsistent observations at both LUX and CTA, which would point to a two-component DM sector.  Such a situation is possible only for $m_\phi \lesssim 300$~GeV and $m_\psi \sim 1$--2~TeV.  Even in this ideal case, measuring the SA coupling $y$ would be difficult.  The best chance would be if the fermion was either lighter or heavier than a wino.  The former might offer evidence of the shifted position of the resonance, while the latter would point to the need of SA or DME to deplete the fermion relic abundance.  In other situations, we would likely only observe one DM particle for some time.  The evidence for the existence of a second state would be difficulty in reconciling the measured DM properties with the observed relic abundance.  However, direct evidence of both DM states from a 100~TeV collider would probably be required to rule out alternative explanations, such as non-thermal production.

\acknowledgments

We thank F. Gao and M. A. Schmidt for valuable discussions. YC and AS were supported by the Australian Research Council.  This work was supported by IBS under the project code, IBS-R018-D1.

\end{document}